\documentclass{INTERSPEECH2023}

\usepackage{subfiles}
\usepackage{algorithm}
\usepackage{algpseudocode}

\usepackage{color, colortbl}
\definecolor{LightGray}{gray}{0.9}

 \interspeechcameraready 
\usepackage{algorithm}

\title{Iterative autoregression: a novel trick to improve your low-latency speech enhancement model}
\name{Pavel Andreev$^{1*}$, Nicholas Babaev$^{1*}$, Ivan Shchekotov$^{123}$, Azat Saginbaev$^1$, Aibek Alanov$^{24}$}
\address{
  $^1$Samsung Research \quad $^2$HSE University \\
  $^3$Skolkovo Institute of Science and Technology \\
  $^4$Artificial Intelligence
Research Institute \quad $^*$equal contribution}
\email{p.andreev@samsung.com, n.babaev@partner.samsung.com}

\begin{document}
\maketitle

\begin{abstract}

Streaming models are an essential component of real-time speech enhancement tools. The streaming regime constrains speech enhancement models to use only a tiny context of future information. As a result, the low-latency streaming setup is generally considered a challenging task and has a significant negative impact on the model's quality. However, the sequential nature of streaming generation offers a natural possibility for autoregression, that is, utilizing previous predictions while making current ones. The conventional method for training autoregressive models is teacher forcing, but its primary drawback lies in the training-inference mismatch that can lead to a substantial degradation in quality. In this study, we propose a straightforward yet effective alternative technique for training autoregressive low-latency speech enhancement models. We demonstrate that the proposed approach leads to stable improvement across diverse architectures and training scenarios.
\end{abstract}

\noindent\textbf{Index Terms}:
speech enhancement; low latency; autoregression.


\section{Introduction}
\label{sec:intro}

The problem of real-time streaming speech processing holds immense practical significance for modern digital hearing aids, acoustically transparent hearing devices, and telecommunication. While the upper limit of undetectable lag for live, real-time processing is a subject of investigation and debate, it is estimated to be around 5-30 milliseconds depending on the application \cite{graetzer2021clarity, stone1999tolerable}. Since speech enhancement tools are usually employed in joint pipelines with other speech processing tools (such as echo cancellation) and within signal transmission channels, requirements for total delay are extremely strict and, for many applications, are hardly met by mainstream speech enhancement solutions that typically rely on algorithmic latency of more than 30-60 ms (by model design) \cite{defossez2020real, hao2021fullsubnet}. Thus, there is a significant demand for research devoted to low-latency (less than 10 ms) speech enhancement models.

The nature of streaming generation follows a sequential pattern that lends itself well to autoregression. The conventional approach for training autoregressive models is through "teacher forcing" \cite{williams1989learning}, whereby the model is presented with past ground-truth samples to predict the subsequent ones during training. During the inference stage, the model utilizes its own samples for autoregressive conditioning (free-running mode) since ground-truth is not available. Teacher forcing is an efficient means of training and convergence as it can be parallelized effectively for convolution-based networks. However, its primary limitation is the mismatch between training and inference which can result in a significant degradation in quality during the test phase. We observe that autoregressive speech enhancement models rely heavily on ground-truth conditioning and are, therefore, particularly vulnerable to training-inference mismatch. The common approaches address this mismatch by utilizing free-running mode during training~\cite{bengio2015scheduled}. However, these methods are typically used for recurrent networks as their application to convolution-based networks substantially slows down training, which hampers practical application.

In this paper, we present a straightforward yet highly efficient algorithm for training autoregressive models that significantly reduces the training-inference mismatch. 
Our approach is based on the iterative substitution of ground-truth conditioning with the model's predictions in teacher-forcing mode. Specifically, we divide the entire training process into $N$ stages, starting with the standard teacher forcing in the initial stage. In the second stage, the forward pass of the model comprises two steps: in the first step, model predicts conditioned on the ground-truth, in the second step, it predicts conditioned on the predictions from the first step. Similarly, at the $n$-th stage, the forward pass consists of $n$ steps, and at each step, the model is conditioned on the predictions from the previous step. We refer to this algorithm as "iterative autoregression" (IA) and demonstrate that it helps to alleviate the training-inference mismatch caused by teacher forcing. Furthermore, we show that autoregressive conditioning offers significant advantages over non-autoregressive baselines across various training losses and neural architectures. Notably, our proposed IA algorithm is versatile and can potentially be applied to training autoregressive models beyond the speech enhancement domain.

\section{Related Work}

\subsection{Low-latency speech enhancement}
Low-latency speech enhancement has recently attracted significant attention from the research community. Many works~\cite{tu2021two, zmolikova2021but} explored time-domain causal neural architectures (e.g., ConvTasNet~\cite{luo2019conv}) for this task, while others~\cite{wang2021deep,wang2022stft,liu2023inplace} utilize time-frequency domain architectures. We limit the scope of this paper to the improvement of time domain architectures, but the proposed method is not bound to a particular domain.

\subsection{Autoregressive models}

Autoregressive models in application to conditional waveform generation are frequently used for neural vocoding~\cite{ordwavenet, kalchbrenner2018efficient, vipperla2020bunched, morrison2021chunked}. WaveNet~\cite{ordwavenet} is a pioneering work proposing a fully-convolutional autoregressive model that produces highly realistic speech samples conditioned on linguistic features. Similarly to this work, we use causal convolutions for autoregressive conditional generation, however, we use a very different type of conditional information (degraded waveform) and generate waveform samples in chunks instead of one by one. CARGAN~\cite{morrison2021chunked} proposes to combine autoregressive conditioning with the power of generative adversarial networks to mitigate artifacts during spectrogram inversion. We also combine autoregressive conditioning with adversarial training, however, we consider a different task and use much smaller chunk sizes (2-16 ms compared to 92 ms employed in \cite{morrison2021chunked}).


\section{Proposed Method}

\begin{figure*}[!h]
  \centering
  \includegraphics[width=0.85\textwidth]{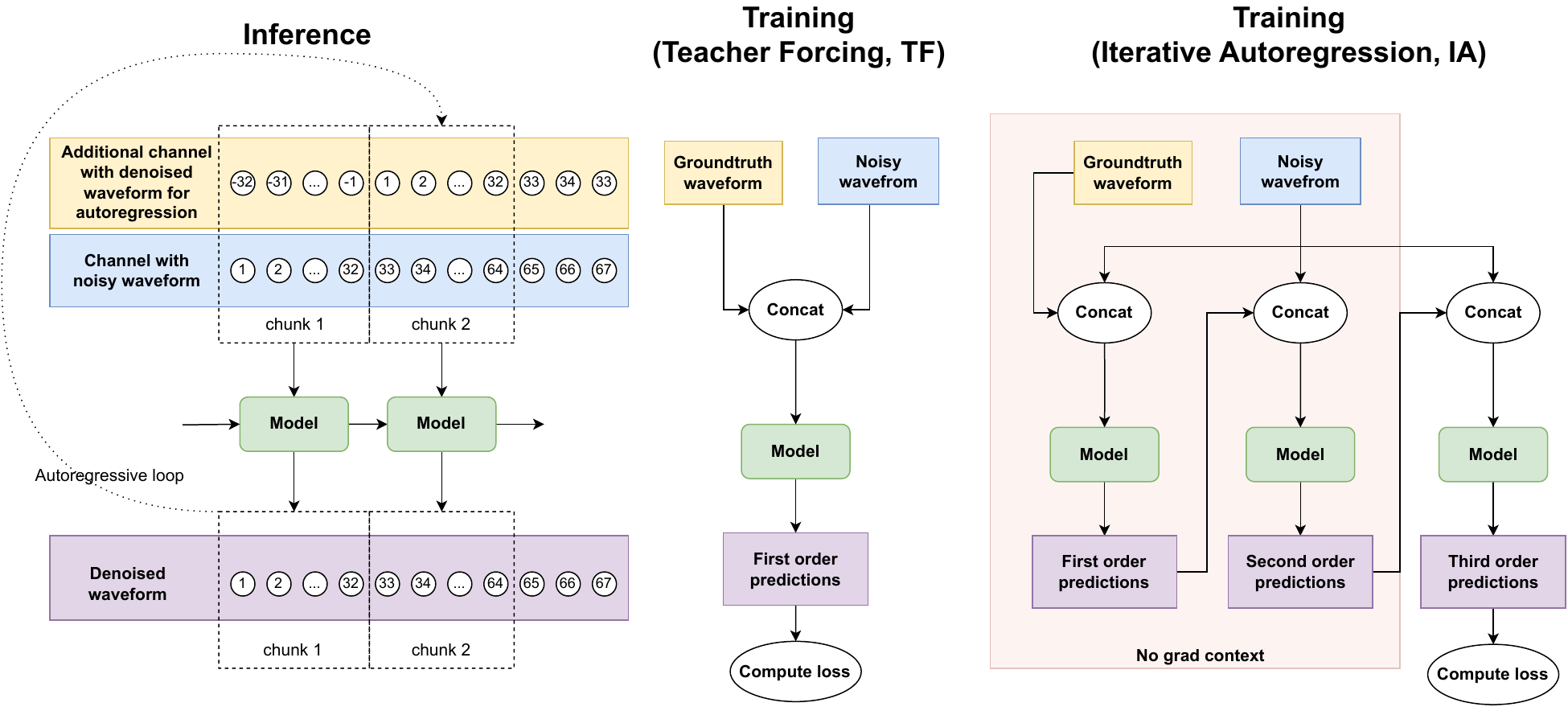}
  \caption{\textit{Left:} illustration of autoregressive conditioning for the model with algorithmic latency 32 timesteps (2 ms at 16 kHz sampling rate). Predicted timesteps from chunk 1 are re-used then making predictions for chunk 2 during inference (free-running mode). \textit{Middle:} Illustration of teacher forcing training. \textit{Right:} Stage 3 of the iterative autoregression training process. The model uses its own predictions to produce predictions of higher orders. We shift ground-truth waveform and predictions before forming a channel with autoregressive conditioning to avoid leakage of future information.}
  \label{fig:llse-inf}
\end{figure*}

\subsection{Limitations of teacher forcing}
The approach of teacher forcing~\cite{williams1989learning} has been widely adopted to train autoregressive models~\cite{morrison2021chunked, ordwavenet, brown2020language}. 

In contrast to the free-running mode, teacher forcing does not require sequential non-parallelizable processing, which is particularly advantageous for convolutional autoregressive models that can be efficiently parallelized during training~\cite{ordwavenet}. Without parallelization, training of such models could take an inordinate amount of time. For instance, for a 2-second audio fragment, autoregressive free-running inference on an NVIDIA v100 GPU takes 1000 times longer than the teacher-forcing inference (forward pass at the training stage) for the WaveNet model, even when using an efficient implementation with activation caching~\cite{paine2016fast}. However, we have observed that the training-inference mismatch caused by teacher forcing can lead to a significant degradation in speech enhancement quality, as shown in Table \ref{tab:maintable}. There are several techniques to alleviate this issue~\cite{bengio2015scheduled, lamb2016professor, dou2020attention}, but all of them rely on the ability to perform autoregressive inference in free-running mode during training, which can complicate the usage of such techniques in practice for convolution-based architectures.

\subsection{Iterative autoregression}
\begin{figure}[!h]
  \centering
  \includegraphics[scale=0.15]{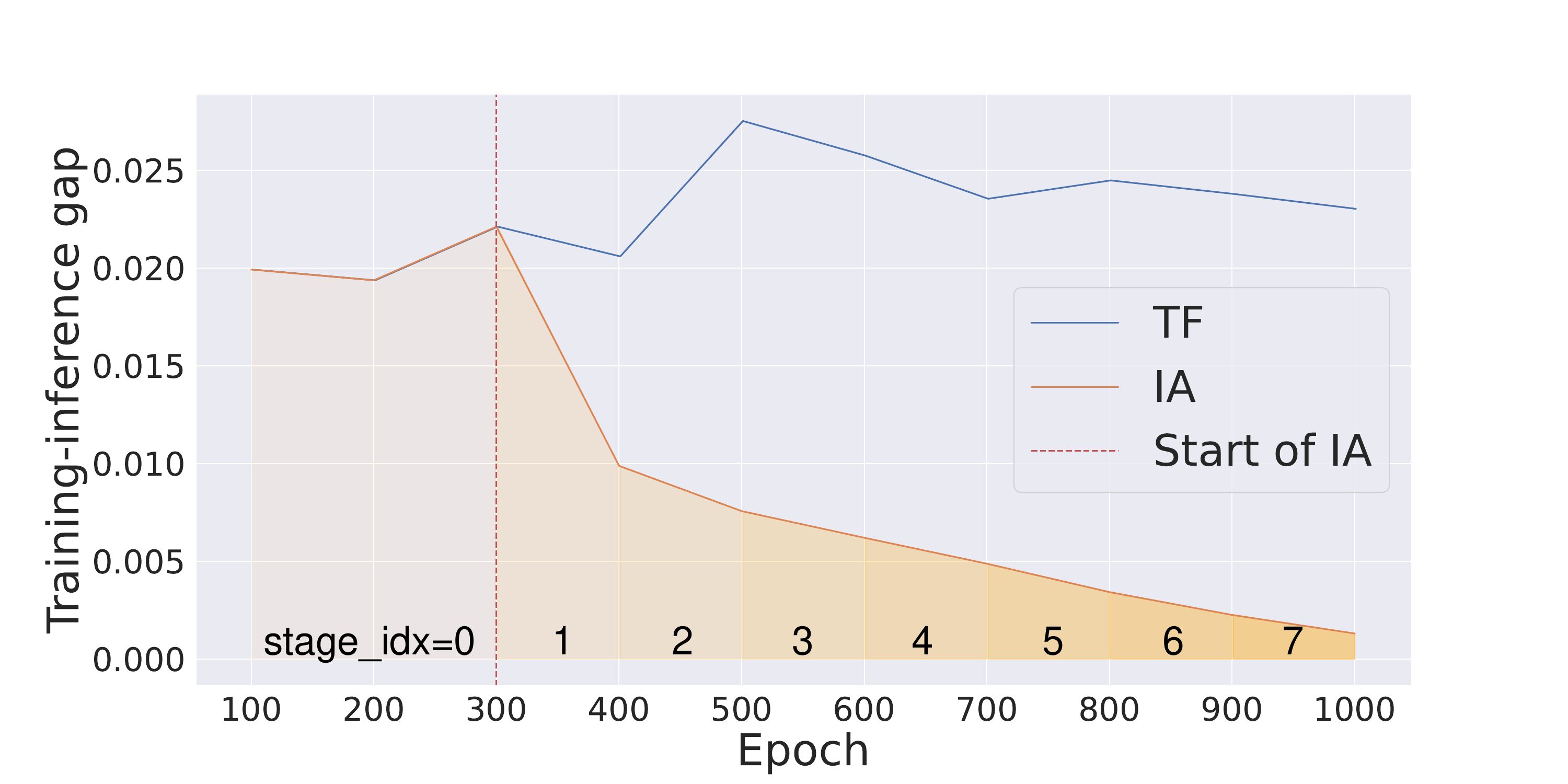}
  \caption{Training-inference mismatch measured as the mean absolute error between model's outputs in inference (free running) and training modes, depending on training epoch. In contrast to teacher forcing (TF), iterative autoregression (IA) causes the output of the training mode to become closer to the output of the inference mode as the training proceeds. Thus, iterative autoregression allows for mitigating training-inference mismatch and improving quality.}
  \label{fig:test-train-diff}
\end{figure}

We have discovered an alternative approach for reducing the discrepancy between training and inference without necessitating the use of a time-consuming free-running mode during training. Our proposal involves the iterative replacement of autoregressive conditioning with the model's predictions in teacher-forcing mode. Initially, the model is trained using the standard teacher-forcing mode, with the autoregressive channel containing the ground-truth waveform. At the next stage, during the forward pass of the model, two steps are involved: (1) the model makes predictions in the teacher-forcing mode, (2) it predicts when the autoregressive channel contains its predictions from the first step. In each subsequent stage of training, the autoregressive input channel comprises the model's predictions as  if they were obtained in the previous stage. Overall, in this training procedure, the model is conditioned on its own predictions in iterative manner. As the training proceeds, we gradually increase the order of predictions for the model to be conditioned on, i.e., increase the number of forwarding passes ("iterations") before computing loss and doing backward passes. Note that we propagate gradient only through the last forward pass. Considering the standard training pipeline which includes forward pass, calculating loss, backpropagation, and weights optimization, the suggested method affects only forward pass, specifically, the forward function of the model. Our method is summarized in Algorithm \ref{alg:iter} and in Figure \ref{fig:llse-inf}. We found that this iterative procedure greatly reduces the mismatch between training and free-running (inference) modes (see Figure \ref{fig:test-train-diff}). Since IA does not require free-running (chunk-by-chunk) predictions during training, each iteration of IA can be efficiently parallelized.

\begin{algorithm}
\caption{Iterative autoregressive training}\label{alg:iter}
\begin{algorithmic}
\Require $f(x, c)$ - model that takes noisy audio $x$ and autoregressive conditioning $c$; $S$ - number of stages; $\mathrm{stage\_iters}$ - list of number of iterations in each stage; $l$ - algorithmic latency of the model in timesteps; $\mathrm{shift}(x, n)$ - function, which removes the last $n$ elements from the last dimension of the tensor $x$ and pads $n$ zeros in the beginning of $x$.
\\
\Procedure{IterativeForwardPass}{$f$, $x$, $c$, $\mathrm{stage\_idx}$}
    \For{$k$ = 1 to $\mathrm{stage\_idx}$ }
    \State \textbf{with NO\_GRAD do}
    \State \ \ \ \ \ \  $c$ $\gets$ $\mathrm{shift}(c, l)$ \algorithmiccomment{Shifting} 
    \State \ \ \ \ \ \  $c$ $ \gets$ $f(x, c)$ \algorithmiccomment{Conditioning} 
    \EndFor
    \State $c$ $ \gets$ $\mathrm{shift}(c, l)$
    \Return $c$
\EndProcedure
\\
\State \textbf{begin}
\For{$\mathrm{stage\_idx}$ = 0 to S}

\For{$i$ = 1 to $\mathrm{stage\_iters[stage\_idx]}$}
\State $(x, y) = \mathrm{sample}\_\mathrm{batch}()$
\State $c$ $\gets$ $ y$
\State $c$ $\gets$  \Call{IterativeForwardPass}{$f$, $x$, $c$, $\mathrm{stage\_idx}$}
\State $\hat{\mathrm{y}}$ $ \gets$ $f(x,c)$
\State $\mathrm{loss}$ $ \gets$ $\mathrm{LossFunc}(y, \hat{\mathrm{y}})$
\State $\mathrm{loss.backward}()$
\State $f\mathrm{.update}()$
\EndFor

\EndFor
\end{algorithmic}
\end{algorithm}

Formally speaking, the waveform generated in the free-running mode is a stationary point of the presented iterative process. Employing notation from Algorithm 1, it is easy to see that if $c$ is a waveform generated in the free-running mode, then $c = \mathrm{IterativeForwardPass}(f, x, c, \mathrm{stage\_idx})$ for any $\mathrm{stage\_idx}$, since the prediction of each next chunk was conditioned on past predictions, i.e., the process resembles the free-running mode. 

Another important insight is that $y^{fr}_{1:n} = \mathrm{IterativeForwardPass}(f, x, c, n)_{2:n+1}$, disregarding of $c$, where $y^{fr}_{1:n}$ are the first $n$ chunks of free-running mode waveform and $\mathrm{IterativeForwardPass}(f, x, c, n)_{2:n}$ are 2 to $n+1$ chunks of conditioining waveform produced by iterative forward pass. Indeed, the first chunk is predicted conditioned on padding for both iterative forward pass and free-running mode. Thus, for $n = 1$ the equation is true. Let the equation be true for $n -1$. Since free-running mode and iterative forward pass were provided with the same conditioning $y^{fr}_{1:n-1}$, their predictions for $n$-th chunk are identical. Thus, by induction $y^{fr}_{1:n} = \mathrm{IterativeForwardPass}(f, x, c, n)_{2:n+1}$ for any $n$. This equality states that iterative autoregression forward pass is guaranteed to converge to the free-running mode waveform (shifted) in a finite number of steps being less or equal to the number of chunks in the waveform. In practice, we found that the process converges much faster than this number. 

\subsection{Model architecture}
Typically, low-latency speech enhancement models are composed of causal neural layers (e.g, uni-directional LSTMs, causal convolutions, causal attention layers, etc.) operating either in time or frequency domains. We propose to directly concatenate the autoregressive conditioning to the model’s input 
(see Figure \ref{fig:llse-inf}). For most of our experiments, we use a simple time domain architecture, which we call WaveUNet+LSTM. It is based on the famous convolutional encoder-decoder UNet architectures~\cite{stoller2018wave} and augmented with a one-directional LSTM layer at the bottleneck to enable a large receptive field to the past time steps similarly to~\cite{defossez2020real}. We use a strided convolutional downsampling layer with kernel size 2 and stride 2 and nearest neighbor upsampling within the UNet structure. The algorithmic latency of this neural network is regulated by the number of downsampling/upsampling layers $K$ and is equal to $2^{K}$ timesteps.

\begin{figure}[!h]
  \centering
  \includegraphics[width=0.45\textwidth]{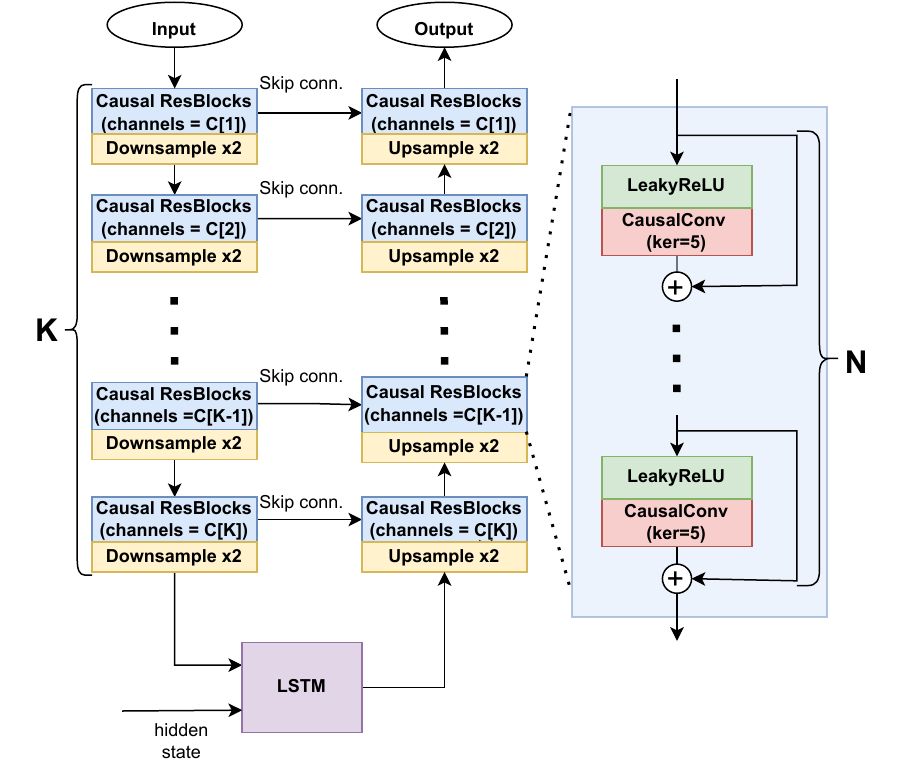}
  \caption{WaveUNet+LSTM architecture. Parameter $K$ regulates the overall depth of the UNet structure, parameter $N$ determines the number of residual blocks within each layer, and array $C$ determines the number of channels on each level of the UNet structure.}
  \label{fig:arch}
\end{figure}
 \vspace{-0.5cm}




\section{Experiments}

\noindent  \textbf{Overall experimental setup} \ \ \
In all our experiments, we consider additive noise as the distortion to be removed from speech recordings.
We conduct a number of experiments to test the proficiency of iterative autoregression in different training scenarios.
For each experimental setting, we train the baseline model without autoregressive conditioning (w/o AR) and the model with autoregressive conditioning (w/ AR). The training conditions and models are identical except AR models are trained with iterative autoregression if not stated otherwise.
The experiments can be divided into 5 settings, depending on the training scenario employed. 
In each training scenario, we change only one training condition (dataset/model architecture/loss/latency) while leaving other parameters as in the base configuration described below.

\noindent \textbf{Metrics} \ \ \  We use state-of-the-art objective speech quality metric UTMOS~\cite{saeki2022utmos}. In our internal experiments, we found this metric has the best correlation among well-known objective metrics (e.g., SI-SDR, DNSMOS, PESQ) with human-assigned MOSes for speech enhancement task. We additionally report conventional metrics DNSMOS~\cite{reddy2022dnsmos} and SI-SDR~\cite{le2019sdr}. To validate the improvement provided by autoregression, we conduct subjective Comparison Category Rating (CCR) tests~\cite{ccr}. 
The referees listened to the output of a baseline (non-autoregressive) model and the corresponding output of an autoregression model and give a score from -3 (if the baseline's output is better) to 3 (if the autoregression's output is better). These tests result in Comparison Mean Opinion Scores (CMOS).

\noindent \textbf{Training hyperparameters} \ \ \ We train all AR models for 1000 epochs and all non-AR models for 2000 epochs (so the training time is approximately the same for the corresponding runs), each epoch includes 1000 batch iterations. The best epoch is selected by maximizing the UTMOS on validation data.  In all experiments, the batch size equals 16, the segment size is set to 2 s, Adam optimizer is used with the learning rate of 0.0002 and betas 0.8 and 0.9. 
We use 8 stages for iterative autoregressive training,  the model is trained for 300 epochs during the initial (teacher forcing) stage and for 100 epochs during all other 7 stages. All experiments were conducted on NVIDIA v100 GPUs, the average experiment duration is around 10 days.
\\

 \vspace{-0.5cm}
\subsection{Experimental series description}
\textbf{Base configuration} \ \ \ 
For the main configuration, we fix the $K$, $N$, and $C$ parameters of the model architecture to 7, 4, and [16, 24, 32, 48, 64, 96, 128], respectively. LSTM width equals 512. This configuration corresponds to 8 ms of algorithmic latency and 6 millions parameters. 
We use the time-domain $L1$ loss function to train the model. 
We employ VoiceBank-DEMAND dataset~\cite{valentini2017noisy} which is a standard benchmark for speech-denoising systems. 
The train set consists of 28 speakers with 4 signal-to-noise ratios (SNR)
(15, 10, 5, and 0 dB) and contains 11572 utterances. The test
set (824 utterances) consists of 2 speakers unseen by the model
during training with 4 SNR (17.5, 12.5, 7.5, and 2.5 dB).
In this base configuration, we also test the proposed iterative autoregression training algorithm against teacher forcing (TF).  It can be seen that autoregressive conditioning leads to much worse results than a non-autoregressive baseline when the model is trained with teacher forcing.  The most characteristic artifacts that we have observed are the regions of silence that  appear in the predicted waveform (we attach samples in supplementary material). The model seems to too heavily rely on ground-truth conditioning, to detect regions of speech and silence. 
\\
\textbf{Different loss functions} \ \ \  
In this series, we test the proficiency of autoregressive conditioning when training with different loss functions. Specifically, we replace $L1$ loss function with adversarial learning losses~\cite{andreev2022hifi++, shchekotov2022ffc} and SI-SNR loss~\cite{luo2019conv}.
One can see that the iterative autoregression improves the results regardless of the loss function employed for training. This result is particularly important in the case of adversarial losses, which do not require exact reconstruction of the ground-truth signal.
\\
\textbf{DNS dataset}  \ \ \  
In this experiment, we use the Deep Noise suppression
(DNS) challenge dataset~\cite{dubey2022icassp} instead of VoiceBank-DEMAND. We synthesized 100 hours of training
data using provided codes and default configuration from Deep Noise Suppression challenge official repository. The models are tested on a hold-out data randomly
selected and excluded from synthesized 100 hours of training
data. 
\\
\textbf{ConvTasNet} \ \ \  
As an alternative architecture, we leveraged the ConvTasNet~\cite{luo2019conv}, a deep
learning framework for end-to-end time-domain speech separation, which can be used as a speech enhancer as well. ConvTasNet consists of a linear encoder, decoder, and a temporal
convolutional network (TCN) consisting of stacked 1-D dilated
convolutional blocks. For experiments, the original architecture implementation was used with the parameters adjusted to match the algorithmic latency to 8 ms and the number of multiply-accumulate operations per second to 2 billion (2 GMAC, the same as in the base configuration).
\\
\textbf{Different latencies}  \ \ \  
To investigate the effect of algorithmic latency on autoregressive improvement, we additionally examine 3 models with latencies of 2 ms, 4 ms, and 16 ms. We tune the parameters of WaveUNet+LSTM architecture to adjust algorithmic latency while leaving the number of multiply-accumulate operations the same as in the base configurations.

\newcommand{\mlcell}[2][p{2cm}c]{%
    \begin{tabular}[#1]{@{}c@{}}#2\end{tabular}
}

\begin{table}[!h]
\caption{\label{tab:maintable} Autoregressive training improves speech enhancement quality in different scenarios. All AR models are trained
with iterative autoregression if not indicated otherwise (TF). } 
\vspace{-0.2cm}
\centering 
\scalebox{0.92}{
\begin{tabular}{c c c c c}
\hline\hline 

\textbf{Experiment} & \textbf{UTMOS} & \textbf{DNSMOS} & \textbf{SISDR} & \textbf{CMOS}  \\ [0.5ex] 

\hline\hline
\multicolumn{5}{c}{\textbf{Base configuration}} \\
\hline\hline
\rowcolor{LightGray} w/o AR & 3.53 & 2.97 & 17.0 &  - \\
\rowcolor{LightGray} w/ AR (TF) & 3.38 & 2.92 & 12.6 &  -0.5\textsuperscript{\scalebox{0.8}{$\pm$ 0.08}}\\
\rowcolor{LightGray} w/ AR  & \textbf{3.61} & \textbf{3.03} & \textbf{18.4} &  0.1\textsuperscript{\scalebox{0.8}{$\pm$ 0.05}}\\ 

\hline\hline
\multicolumn{5}{c}{\textbf{Different losses }}\\
\hline\hline
w/o AR (adv.) & 3.68 & 3.02 & 15.2 & -\\
w/ AR (adv.) & \textbf{3.74} & \textbf{3.04} & \textbf{15.3}  & 0.12\textsuperscript{\scalebox{0.8}{$\pm$ 0.04}}\\ 

\hline\hline
\rowcolor{LightGray}  w/o AR (si-snr) & 3.51 & 2.95 & 17.0 & - \\
\rowcolor{LightGray} w/ AR (si-snr) & \textbf{3.57} & \textbf{2.96} & \textbf{17.8} & 0.13\textsuperscript{\scalebox{0.8}{$\pm$ 0.05}}\\ 

\hline\hline
\multicolumn{5}{c}{\textbf{DNS dataset}} \\
\hline\hline
w/o AR  & 2.42 & 2.98 & 14.5 &  -\\
w/ AR  & \textbf{2.47} & \textbf{3.03} & \textbf{14.6} & 0.1\textsuperscript{\scalebox{0.8}{$\pm$ 0.05}}\\ 

\hline\hline
\multicolumn{5}{c}{\textbf{ConvTasNet architecture}}\\
\hline\hline
\rowcolor{LightGray} w/o AR  & 3.08 & 2.86 & 15.3 &  -\\
\rowcolor{LightGray} w/ AR & \textbf{3.33} & \textbf{2.99} & \textbf{15.8} &  0.52\textsuperscript{\scalebox{0.8}{$\pm$ 0.06}}\\ 

\hline\hline
\multicolumn{5}{c}{\textbf{Different latencies}} \\
\hline\hline
w/o AR (2 ms) & 3.47 & 2.94 & 17.1  & -\\
w/ AR (2 ms) & \textbf{3.55} & \textbf{2.98} & \textbf{18.3} & 0.04\textsuperscript{\scalebox{0.8}{$\pm$ 0.04}}\\ 

\hline	
\rowcolor{LightGray} w/o AR (4 ms) & 3.5 & 2.96 & 17.2 &  - \\
\rowcolor{LightGray} w/ AR (4 ms) & \textbf{3.59} & \textbf{3.02} & \textbf{18.6} & 0.16\textsuperscript{\scalebox{0.8}{$\pm$ 0.05}}\\ 

\hline	
w/o AR (16 ms) & 3.57 & 2.99 & 17.3 & - \\
w/ AR (16 ms) & \textbf{3.64} & \textbf{3.02} & \textbf{18.6} &  0.09\textsuperscript{\scalebox{0.8}{$\pm$ 0.04}}\\ 

\hline\hline
\end{tabular}} 
\label{table:maintable}

\vspace{-0.6cm}
\end{table}

\section{Conclusion}

The proposed method of iterative autoregressive training allows for improving the quality of streaming speech enhancement models in all studied scenarios. Furthermore, it dramatically outperforms the conventional teacher forcing method, which fails to provide any improvements over the non-autoregressive baseline due to a high training-inference mismatch. We believe that the presented technique provides a practical alternative to teacher forcing and takes an important step toward improving streaming models by means of autoregression.

\section{Acknowledgements}
This work was supported by Samsung Research. Aibek Alanov
was supported by the grant provided by the Analytical Center
for the Government of the Russian Federation (ACRF) in accor-
dance with the agreement No. 000000D730321P5Q0002 and
the agreement with HSE University No. 70-2021-00139.

\bibliographystyle{IEEEtran}
\bibliography{A}

\end{document}